\documentclass{webofc}
\usepackage[varg]{txfonts}   
\usepackage{graphicx}
\usepackage{subcaption}
\usepackage[labelfont=bf]{caption}
\usepackage{float}
\usepackage{hyperref}
\hypersetup{colorlinks=true,citecolor=blue,urlcolor=blue,linkcolor=blue}
\newcommand{\uproman}[1]{\uppercase\expandafter{\romannumeral#1}}

\begin{document}
\title{\boldmath Spin-Density Matrix Elements in the Reaction
$\gamma p\rightarrow\rho^-\Delta^{++}$ at GlueX}

\author{
    \firstname{Niklas} \lastname{Herrmann}\inst{1}
    \fnsep\thanks{\email{herrmann@jlab.org}}
    \and
    \firstname{Farah} \lastname{Afzal}\inst{1}
    \firstname{} \lastname{for the GlueX Collaboration}
}
\institute{Institute of Experimental Physics I,
  Ruhr University Bochum, Germany}

\abstract{The GlueX experiment, located in Hall~D at Jefferson Lab, aims to map the full spectrum of light-quark mesons with an emphasis on hybrid mesons, states in which the gluonic field is excited and contributes to the quantum numbers of the meson. 
The search for exotic hybrid mesons requires a detailed understanding of the underlying production mechanisms. Polarization observables, such as the spin-density matrix elements (SDMEs), are an important experimental tool to investigate these mechanisms.
The measurement of the unpolarized and polarized SDMEs in the photoproduction of $\rho^- \Delta^{++}$ using GlueX data  is reported.
  In this analysis, for the first time, the correct full formalism is applied in which the upper mesonic and lower baryonic systems are described by a joint spin-density matrix. 
A factorized model yields results consistent with those of the full model, suggesting that this simplified description may also be suitable for other reactions involving meson–baryon final states. 
}

\maketitle
\section{Introduction}
\label{intro}
 
Compared to the quark model in which mesons are described as simple quark-antiquark states, QCD allows for additional states, including hybrid mesons in which the gluonic field is excited and contributes to the meson quantum numbers. Currently, one of the best experimental candidates for a hybrid meson is the
$\pi_1(1600)$, which has been observed, for example, in COMPASS pion-production data \cite{alexeev_exotic_2022}.\newline
In contrast to ongoing searches at other facilities, the GlueX experiment uses photoproduction, where t-channel Regge-exchanges with different quantum numbers are accessible, as a complementary production process to search for hybrid mesons. To this end, a linearly polarized photon beam is incident on a liquid hydrogen target.  Important tools in the search for hybrid mesons are polarization observables, like spin-density matrix elements (SDMEs) as they are sensitive to the production mechanism. \newline
In this analysis, we determine the SDMEs in the reaction $\gamma p \rightarrow\rho^- \Delta^{++}\rightarrow\pi^- \pi^0\pi^{+}p$ by using the large GlueX-\uproman{1} dataset.
The $\pi^-\pi^0$ system gives access to the $1^{--}$ isovectors like the $\rho^-(770)$ which is the focus of our analysis. In the past, the SDMEs of $\rho^0(770)$ with a recoiling proton \cite{adhikari_measurement_2023} and the SDMEs of $\Delta^{++}(1232)$ with a $\pi^-$ of the upper vertex \cite{afzal_measurement_2025} have been studied by GlueX. The reaction studied in our analysis now focuses on $\rho^-$-photoproduction recoiling against a $\Delta^{++}$. For the first time, the combined treatment of the upper vertex with the meson and the lower vertex with the baryon is conducted, taking the correlations between the $\rho^-$ and $\Delta^{++}$ decays into account. The unpolarized SDMEs of $\rho^-(770)$ in this reaction are compared with early, low-precision measurements by SLAC \cite{abramson_rhopm_1976}.
\section{Experimental setup}
The GlueX experiment \cite{adhikari_gluex_2021} is located in Hall D at Jefferson Lab and uses the 12 GeV electron beam of the Continuous Electron Beam Accelerator Facility (CEBAF). The electron beam scatters off a diamond radiator. Through coherent bremsstrahlung, a linearly polarized photon beam is produced which is incident on a liquid hydrogen target. The polarization is described by the polarization vector: 
$\vec{P}_{\gamma}=P_{\gamma}\left(-\cos(2\Phi),-\sin(2\Phi),0\right)
$, where $\Phi$ denotes the polarization angle, the angle between the reaction plane and the polarization vector, and $P_{\gamma}$ is the degree of polarization. A high polarization of up to $P_{\gamma} \backsim 40\,\%$ is reached in the coherent peak, in the range from $8.2\,\mathrm{GeV} \text{ to } 8.8\,\mathrm{GeV}$. The liquid hydrogen target is surrounded by a spectrometer comprising tracking detectors, calorimeters, and a time-of-flight detector for particle identification.
The spectrometer has the capability of reconstructing both charged and neutral pions and therefore enables GlueX to measure all four possible $\pi\pi$ final states. 
\section{Event selection}
\begin{figure}[h]
    \centering

    \includegraphics[
        width=0.40\textwidth
    ]{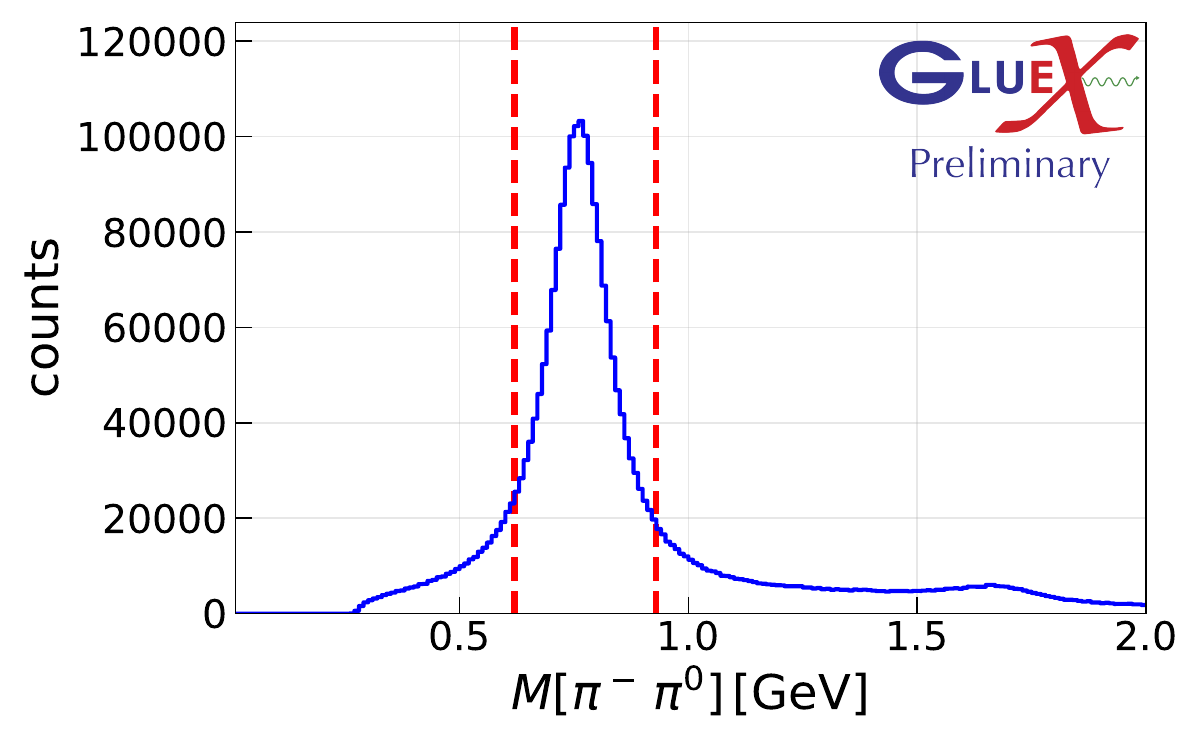}
    \hspace{0.02\textwidth}
    \includegraphics[
        width=0.40\textwidth
    ]{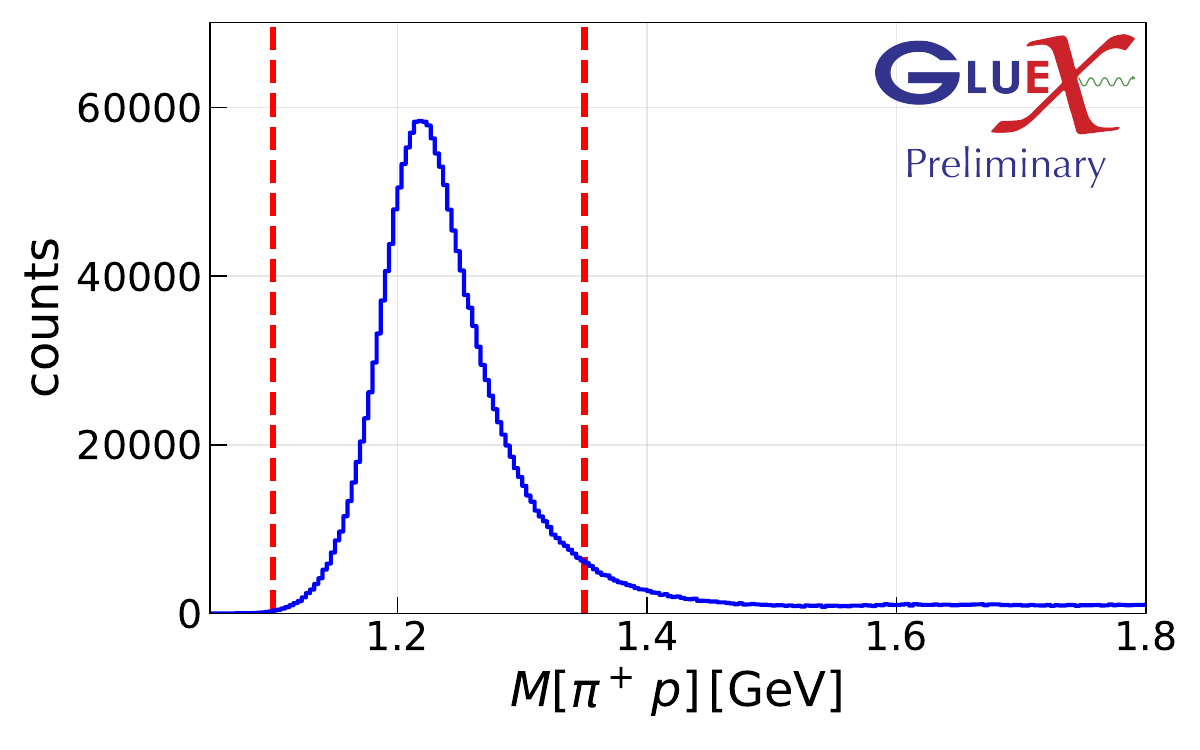}

    \caption{Invariant-mass distributions of $\pi^-\pi^0$ (left) and
    $\pi^+p$ (right) after the event selection. The red lines indicate
    the selection windows.}
    \label{fig:mass}
\end{figure}
\noindent
Events with the exclusive final state $\pi^- \pi^0\pi^{+}p$ are selected and a kinematic fit requiring a common vertex for all charged particles and energy and momentum conservation. The $\pi^0$ is reconstructed via its decay into two photons. Events with a $\chi^2/NDF<5$ for the reaction hypothesis $\gamma p \rightarrow \pi^-\pi^0 \pi^+p$ are retained.  The $\rho^-(770)$ and $\Delta^{++}(1232)$ are selected by restricting the invariant mass ranges of $\pi^0 \pi^-$ and $\pi^+p$ to regions around the resonances, respectively, as depicted in Fig \ref{fig:mass}. Additional requirements on reconstruction quality, beam energy, missing mass, unused tracks and showers, and selected invariant masses are imposed to suppress background.  In total, 1.8 million events of GlueX-\uproman{1} pass the selection criteria, corresponding to approximately 1000 times the number of events used in the SLAC study \cite{abramson_rhopm_1976}.  
The background from events with different topologies is lower than  $2\,\%$, while the combinatorial background is modeled with uniformly distributed phase space events in the fit and varies from  $1\,\%$ to  $14\,\%$ depending on the momentum transfer squared $t$.

\section{Extraction of SDMEs}
\begin{figure}[h]
\centering
\includegraphics[width=13cm,clip]{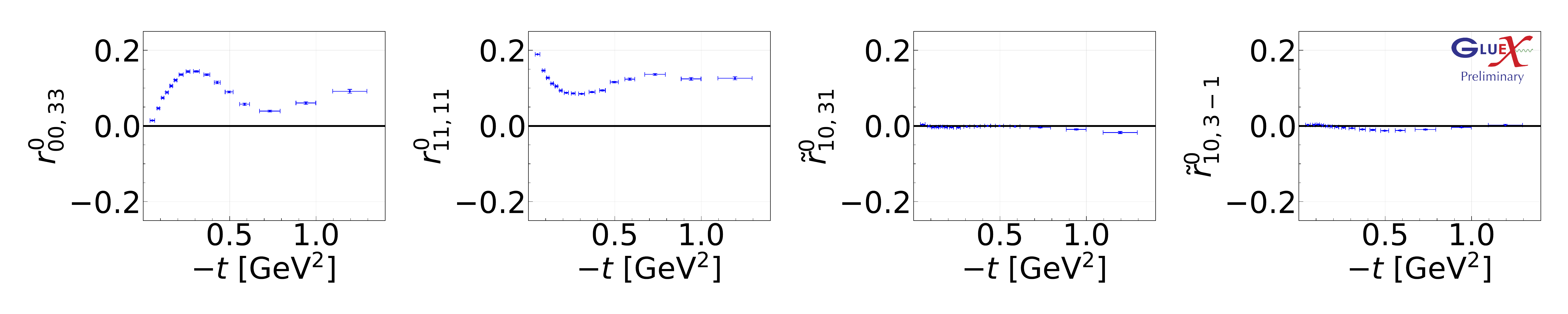}
\caption{Four representative unpolarized double SDMEs ($\alpha=0$) extracted with the full model. The vertical error bars only show the statistical uncertainty.}
\label{doubleSDMEs}       
\end{figure}

\begin{figure}[h]
    \centering

    \begin{subfigure}[t]{0.49\textwidth}
        \vfill
        \centering
        \includegraphics[
            width=\linewidth,
            clip
        ]{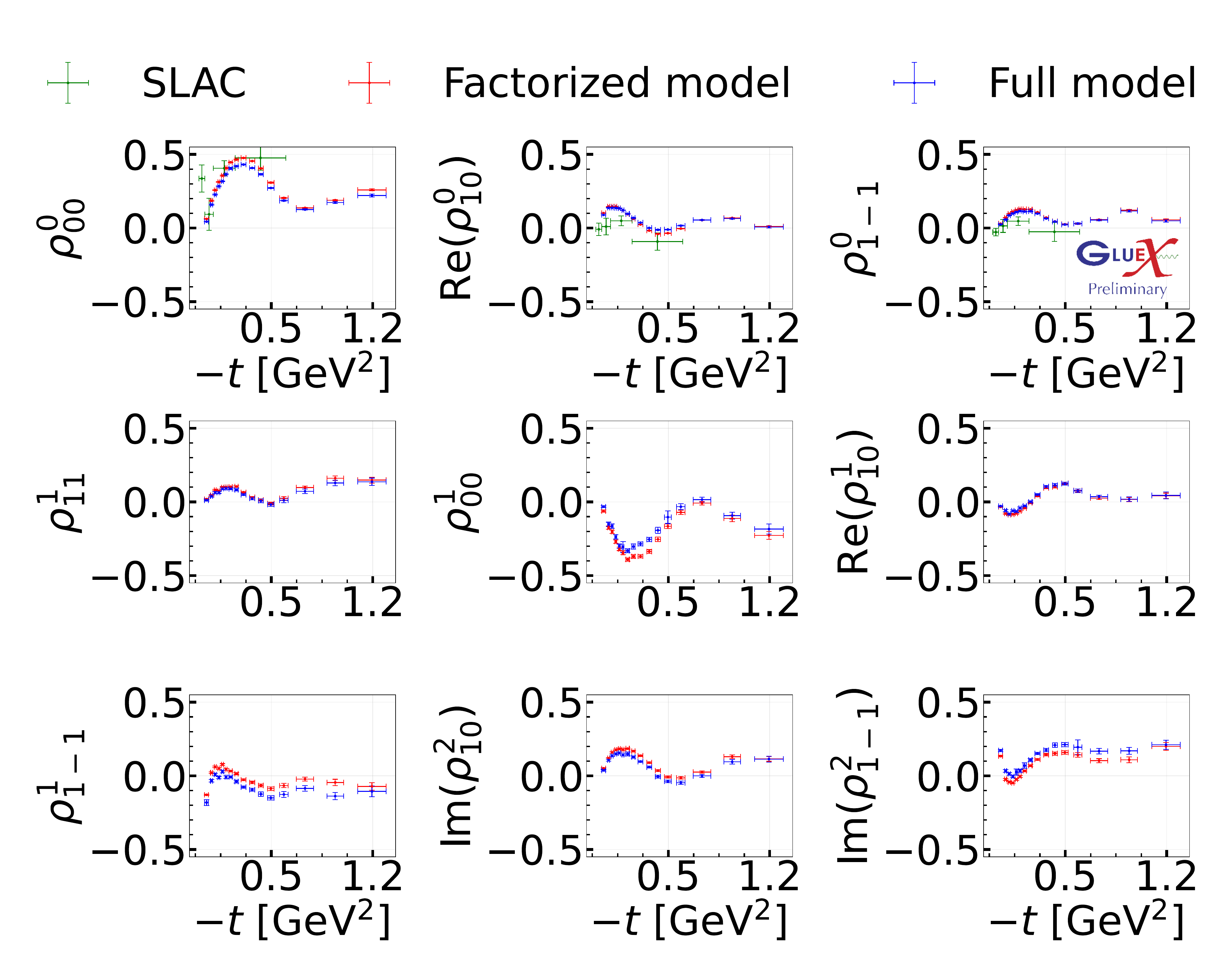}
        \caption{}
        \label{fig:rho_reduced}
    \end{subfigure}
    \hfill
    \begin{minipage}[t]{0.49\textwidth}
        \vfill
        \centering

        \begin{subfigure}[t]{\linewidth}
            \centering
            \includegraphics[
                width=\linewidth,
                clip
            ]{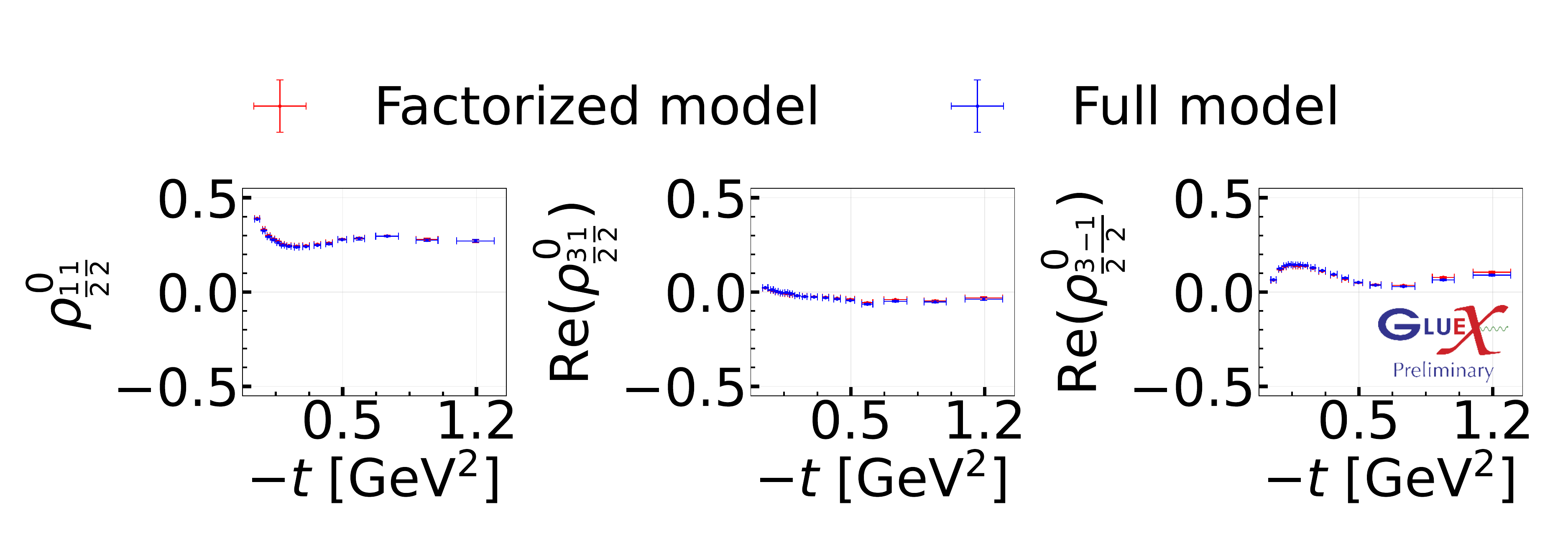}
            \caption{}
            \label{fig:delta_reduced}
        \end{subfigure}

        \begin{subfigure}[t]{\linewidth}
            \centering
            \includegraphics[
                width=\linewidth,clip
            ]{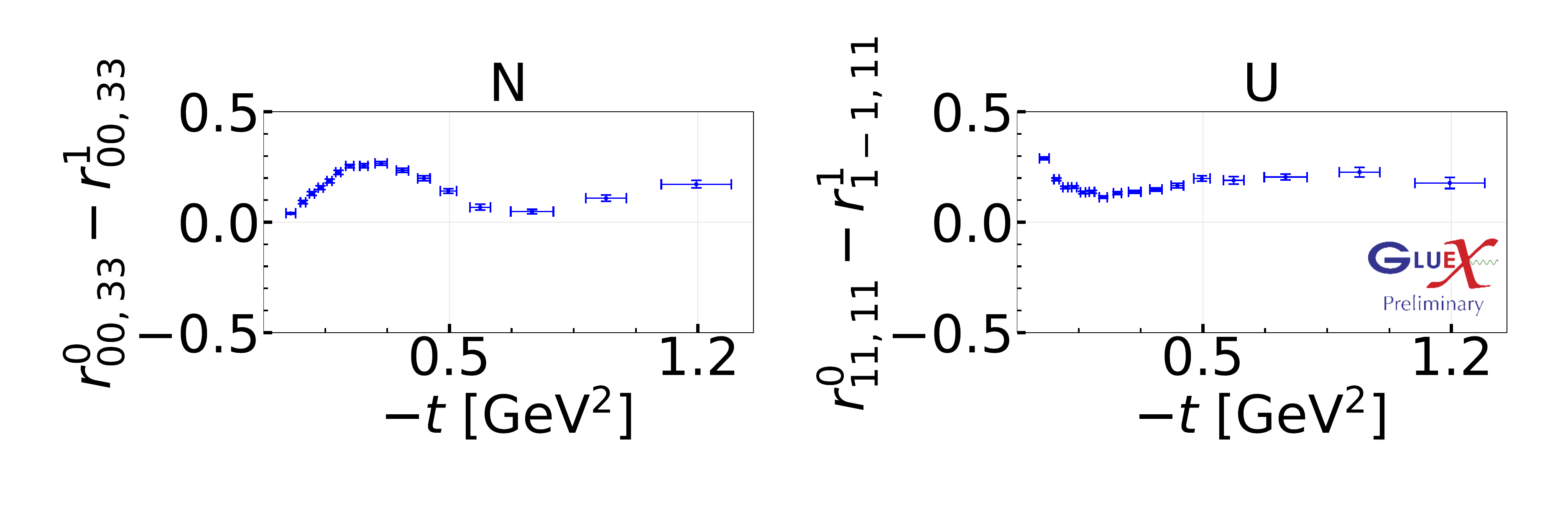}
            \caption{}
            \label{fig:naturality}
        \end{subfigure}
    \end{minipage}

    \caption{Preliminary GlueX results for the SDMEs of the $\rho^-$ (a) and the
    $\Delta^{++}$ (b), obtained using the full and factorized models and compared
    to the SLAC results. Panel (c) shows two representative linear combinations
    of the fit parameters (full model) that correspond to the purely
    natural-parity (N) or purely unnatural-parity (U) contributions.}
\end{figure}

\paragraph{Full model}
\label{Fullmodel}
The photon polarization vector determines the corresponding photon density matrix, allowing the density matrix of the meson-baryon system  to be decomposed into an unpolarized component ($\alpha=0$) and two linearly polarized components ($\alpha=1,2$). In matrix form, they are written as double-indexed SDMEs $\rho^\alpha_{mm'\lambda_{\Delta}\lambda'_{\Delta}}$, where the helicity of $\rho^-$ is denoted by $m,m'$ and the helicity of $\Delta^{++}$ is labeled $\lambda_{\Delta},\lambda_{\Delta}'$.
The full formalism and the derivation of the intensities as function of the SDMEs are described in \cite{mathieu_meson_nodate}. The full model intensity is decomposed in two polarized and one unpolarized components:\begin{equation}
I(\Omega^{\pi^-},\Omega^{p},\Phi)
=
I^{0}(\Omega^{\pi^-},\Omega^{p})
-
P_{\gamma}I^{1}(\Omega^{\pi^-},\Omega^{p})\cos(2\Phi)
-P_{\gamma}
I^{2}(\Omega^{\pi^-},\Omega^{p})\sin(2\Phi)
\end{equation} The unpolarized full model intensity for example is given as \cite{mathieu_meson_nodate}:
\begin{equation}
\begin{aligned}
I^0(\Omega^{\pi^-},\Omega^p)
={}&
\cos^2\theta^{\pi^-}
\left[
W_{00}^{0}(\Omega^p)
-
\overline{W}_{00}^{0}(\Omega^p)
\right]
+
\sin^2\theta^{\pi^-}
\left[
W_{11}^{0}(\Omega^p)
-
\overline{W}_{11}^{0}(\Omega^p)
\right]
\\[2mm]
&-
\sin^2\theta^{\pi^-}
\left(
\cos(2\phi^{\pi^-})
\left[
W_{1-1}^{0}(\Omega^p)
-
\overline{W}_{1-1}^{0}(\Omega^p)
\right]
+
\sin(2\phi^{\pi^-})\,
\widetilde{W}_{1-1}^{0}(\Omega^p)
\right)
\\[2mm]
&-
\sqrt{2}\sin(2\theta^{\pi^-})
\left(
\cos\phi^{\pi^-}
\left[
W_{10}^{0}(\Omega^p)
-
\overline{W}_{10}^{0}(\Omega^p)
\right]
+
\sin\phi^{\pi^-}\,
\widetilde{W}_{10}^{0}(\Omega^p)
\right).
\end{aligned}
\label{eq:unpolarized_intensity}
\end{equation}

with 
\begin{equation}
\begin{aligned}
W_{mm'}^{\alpha}(\Omega)
&=
r^{\alpha}_{mm'33}\sin^2\theta
+
r^{\alpha}_{mm'11}
\left(\frac{1}{3}+\cos^2\theta\right),
\\
\overline{W}_{mm'}^{\alpha}(\Omega)
&=
\frac{2}{\sqrt{3}}
\left[
r^{\alpha}_{mm'31}\cos\phi\,\sin(2\theta)
+
r^{\alpha}_{mm'3{-1}}\cos(2\phi)\,\sin^2\theta
\right],
\\
\tilde{W}_{mm'}^{\alpha}(\Omega)
&=
\frac{2}{\sqrt{3}}
\left[
\tilde{r}^{\alpha}_{mm'31}\sin\phi\,\sin(2\theta)
+
\tilde{r}^{\alpha}_{mm'3{-1}}\sin(2\phi)\,\sin^2\theta
\right].
\end{aligned}
\label{eq:angular_functions}
\end{equation}
The fit parameters $r$ and $\tilde{r}$ are linear combinations of the double SDMEs. If the production amplitude factorizes, the $\tilde{r}$ fit parameters are expected to be zero \cite{mathieu_meson_nodate}.  The joint density matrix contains 432 SDMEs. However, after imposing normalization, hermiticity, and parity constraints, only 55 independent linear combinations enter the intensity and are used as fit parameters.
To determine the SDMEs as a function of the momentum transfer squared, the considered range $0.025\,\mathrm{GeV}^2<-t< 1.4\,\mathrm{GeV}^2$ is divided into 16 bins. An extended unbinned maximum-likelihood fit is used to extract the SDMEs, with the detector acceptance evaluated using simulated phase-space events.
 The angles $\Omega_{\pi^-}=(\phi^{\pi^-},\theta^{\pi^-}$) and $\Omega_p=(\phi^{p},\theta^{p})$ are calculated in the respective rest frames of the $\rho^-$ and
$\Delta^{++}$ using the Gottfried--Jackson (GJ) coordinate system, following the conventions used in ~\cite{adhikari_measurement_2023,afzal_measurement_2025}.
\noindent Figure \ref{doubleSDMEs} shows the results for the unpolarized SDMEs ($\alpha = 0$). The vertical error bars only represent the statistical uncertainty. All $\tilde{r}$ are close to zero, which supports the factorization hypothesis. This trend also remains for $\alpha=1,2$. 
Certain linear combinations of the fit parameters separate natural-parity from unnatural-parity exchange.
Fig. \ref{fig:naturality} shows a representative linear combination of the fit parameters for each of the natural-parity and unnatural-parity contributions. Natural-parity exchange dominates for $0.07\,\mathrm{GeV}^2 < -t < 0.40\,\mathrm{GeV}^2$, while unnatural-parity exchange dominates at larger $-t$ and at very small $-t < 0.07\,\mathrm{GeV}^2$.
\paragraph{Factorized model}
Since the SDMEs extracted with the full model support the factorization hypothesis, they motivate the use of a simpler factorized fit model. This model is again described in \cite{mathieu_meson_nodate}. Under the factorization hypothesis, only the $\rho^-$ subsystem is sensitive to the photon polarization. Consequently, the total intensity is written as the product of the standard $\rho^-$ intensity~\cite{adhikari_measurement_2023,schilling_analysis_1970} and the unpolarized $\Delta^{++}$ intensity~\cite{afzal_measurement_2025}.
The assumption of factorization reduces the number of fit parameters from 55 to only 12 fit parameters. To compare the results of the full model with the results of the factorized model the SDMEs are calculated by taking the trace of either the $\Delta^{++}$ or the $\rho^{-}$ subsystem. For example, the SDMEs of $\rho^-$ are given by $\rho^{\alpha,\rho^-}_{mm'}
=
\left(
\operatorname{Tr}_{\Delta^{++}}
\rho^\alpha
\right)_{mm'}
=
\sum_{\lambda_\Delta}
\rho^\alpha_{mm'\lambda_\Delta\lambda_\Delta}$.
The $\rho^-$ and $\Delta^{++}$ SDMEs obtained with the full and factorized models are in good agreement (Fig.~\ref{fig:rho_reduced} and~\ref{fig:delta_reduced}). In addition, the unpolarized SDMEs of $\rho^-$ determined with the full model agree with the SLAC measurement. Compared with the SLAC measurement, our data set contains approximately three orders of magnitude more events, allowing for a finer $t$-binning and substantially improved statistical precision.

\section{Summary and outlook}
In this contribution, we have presented the results for the SDMEs in the reaction $\gamma p \rightarrow\rho^- \Delta^{++}\rightarrow\pi^- \pi^0\pi^{+}p$. For the first time, double SDMEs describing both the mesonic system and the baryonic system have been successfully extracted. Remarkably, the factorized model, requiring only 12 fit parameters, yields results consistent with those of the full model involving 55 parameters. This substantial reduction in complexity is promising for studies of other meson-baryon final states, particularly  $\pi\pi\pi\Delta^{++}$, which provides access to the spin-exotic $\pi_1(1600)$. In the future, the systematic uncertainties will be determined.
\section{Acknowledgments}
This work is supported by the German Research Foundation (DFG) under project number 537210888. We thank the Joint Physics Analysis Center (JPAC), particularly Vincent Mathieu and Vanamali Shastry, for valuable input.
GlueX acknowledges the
support of several funding agencies and computing facilities (\url{http://gluex.org/thanks}).

\bibliography{Meson2026}

\end{document}